\newcommand{\Tr}{\mathop{\mathrm{Tr}} \nolimits}
\newcommand{\rank}{\mathop{\mathrm{rank}} \nolimits}
\newcommand{\LG}[2]{\mathrm{LG}^{#1}_{#2}}

\newcommand{\sugg}[1]{{#1}}
\documentclass[prl,superscriptaddress,twocolumn,showpacs]{revtex4}
\usepackage{amsfonts,amssymb,amsbsy,bm,graphicx,amsmath,times,color}

\begin{document}

\title{Full tomography from compatible measurements}

\author{J. \v{R}eh\'{a}\v{c}ek}
\affiliation{Department of Optics,
Palacky University, 17. listopadu
50, 772 00 Olomouc, Czech Republic}

\author{Z. Hradil}
\affiliation{Department of Optics,
Palacky University, 17. listopadu 50,
772 00 Olomouc, Czech Republic}

\author{Z. Bouchal}
\affiliation{Department of Optics,
Palacky University, 17. listopadu
50, 772 00 Olomouc, Czech Republic}

\author{R. \v{C}elechovsk\'y}
\affiliation{Department of Optics,
Palacky University, 17. listopadu
50, 772 00 Olomouc, Czech Republic}

\author{I. Rigas}
\affiliation{Departamento de \'Optica,
Facultad de F\'{\i}sica,
Universidad Complutense, 28040~Madrid, Spain}

\author{L. L. S\'{a}nchez-Soto}
\affiliation{Departamento de \'Optica,
Facultad de F\'{\i}sica,
Universidad Complutense, 28040~Madrid, Spain}

\begin{abstract}
  We put forward a reconstruction scheme prompted by the relation
  between a von Neumann measurement and the corresponding
  informationally complete measurement induced in a relevant
  reconstruction subspace.  This method is specially suited for the
  full tomography of complex quantum systems, where the intricacies of
  the detection part of the experiment can be greatly reduced provided
  some prior information is available. In broader terms this shows the
  importance of \sugg{this often-disregarded} prior information in
  quantum theory.  The proposed technique is illustrated with an
  experimental tomography of photonic vortices of moderate dimension.
\end{abstract}

\pacs{03.65.Wj, 03.65.Ta, 42.50.Tx}

\maketitle

\textit{Introduction.}  
The quantum state is a mathematical object that encodes complete
information about a \sugg{system~\cite{Peres}}: once it is known, the
outcomes of any possible measurement can be predicted.  Apart from
fundamental reasons, acquiring the system state is invaluable for
verifying and optimizing experimental setups. For instance, in some
protocols of quantum key distribution, the knowledge of the entangled
state distributed between the parties greatly limits the ability of a
third party to eavesdrop on the communication channel~\cite{eve}.

The reconstruction of the unknown state from a suitable set of
measurements is called quantum \sugg{tomography~\cite{lnp}}.  
Over the past years, this technique has evolved from the first
theoretical~\cite{vogel} and experimental~\cite{raymer} concepts to a
widely acknowledged and fairly standard method \sugg{extensively used 
for} both discrete~\cite{james,thew} and 
\sugg{continuous~\cite{lvovsky} variables}.

In this work, we focus on \sugg{measurement strategies for the tomographic
reconstruction, leaving aside data post-processing issues}. In practice, a
sufficient number of independent observations must be included in the
set of measurements so that all physical aspects of the measured
system are addressed.  When dealing with complicated systems,
such measurements may be difficult to implement in the laboratory due
to various physical and technical limitations on the available
controlled interactions between the system and the meter.

The goal of this Letter is to present a method of generating a
tomographically complete measurement set from a simple von Neumann
measurement that is readily implemented in the laboratory. Obviously,
a von Neumann measurement is not complete, as all the measured
projections are compatible and hence provide information only about
the same \sugg{aspects}. However, as we shall show here, things are
radically different when only a part of the full Hilbert space is of
\sugg{interest: In} this subspace, even a simple von Neumann
projection may become informationally complete.  This should not be
taken as an \sugg{approximation, in the sense that some accuracy is
  traded for experimental feasibility}.  First of all, the energy of
any system is always bounded, so one can restrict the attention to the
subspace spanned by low-energy states. Second, due to the finite
resources, all quantum systems are \textit{de facto} discrete and may
be represented by a relatively small number of parameters. In that
case, there is no necessity of sophisticated measurements that are
informationally complete in the original large Hilbert space: since
only a small subset is accessible, even much simpler observations are
able to supply the information needed.  This is the main idea behind
the present contribution.

\textit{Quantum tomography.}  Let us consider a density matrix
$\varrho$ describing a $d$-dimensional quantum system. A convenient
representation of $\varrho$ can be obtained with the help of a
traceless Hermitian operator basis $\{\Gamma_i\}$, satisfying $\Tr
(\Gamma_i)= 0$ and $\Tr (\Gamma_i \Gamma_j) = \delta_{ij}$~\cite{sun}:
\begin{equation}
  \label{rhodecomp}
  \varrho=\frac{1}{d} + \sum_{i=1}^{d^2-1} a_i \Gamma_i \, ,
\end{equation}
where $\{ a_{i} \}$ are real numbers. The set $\{\Gamma_i\}$
coincides with the orthogonal generators of SU($d$), which is the
associated symmetry algebra.

In general, the measurements performed on the system are described by
positive operator-valued measures (POVMs), which are a set of
operators $\{ \Pi_j \}$ (with $\Pi_j\ge 0$ and $\sum_j \Pi_j=
\openone$), such that each POVM element represents a single output
channel of the \sugg{measuring apparatus}.  The probability of
detecting the $j$th output is given by a generalized projection
postulate $p_j=\Tr (\varrho \Pi_j)$.

By decomposing the POVM elements in the same basis
$\{\Gamma_i\}$, we get
\begin{equation}
  \label{PIdecomp}
  \Pi_j = b_j + \sum_{i=1}^{d^2-1} c_{ji} \Gamma_i \, ,
\end{equation}
where $\{ b_{j} \}$ are again known real numbers and
$\mathbf{C} = \{c_{ji}\}$ is a real matrix.

\textit{Informational completeness.} 
A set of measurements will be called informationally complete if any
quantum state $\varrho$ is unambiguously assigned to the corresponding
theoretical probabilities $p_j$. Since the projection postulate 
can be rewritten as
\begin{equation}
  p_j - b_j = \sum_i c_{ji} a_i \, , 
\end{equation} 
informational completeness requires the matrix $\mathbf{C}$ to have at
least $d^2-1$ linearly independent rows. Numerically, this can
be easily verified by calculating the rank of $\mathbf{C}$, given
by the number of nonzero singular values. These are readily computed
from the singular value decomposition of $\mathbf{C}$.  Thus, a set of
measurements is informationally complete provided
\begin{equation}
  \label{condition} 
  \rank \mathbf{C} \ge d^2 - 1 \, .
\end{equation}

For example, a light mode can be \sugg{treated} as a harmonic
oscillator. The eigenstates of the rotated quadrature operators
$Q(\theta) = x \cos \theta + p \sin \theta$ comprise an
informationally complete POVM.  Naturally, only a finite set of
projections can be done, so that a truncation of the original
infinite-dimensional Hilbert space is
necessary~\cite{grangier,polzik}.  In consequence, consider a von
Neumann projection defined in the infinite-dimensional space
$\mathcal{H}$: $ \sum_{k=0}^\infty |k \rangle \langle k | = \openone$,
where $| k \rangle$ is an orthonormal basis.  Experimentally such
measurements do not pose any difficulty: all that has to be done is to
determine the spectrum of a single observable.  Nevertheless, this
simple von Neumann measurement is not informationally complete in
$\mathcal{H}$, for all the observations are in this case mutually
compatible and consequently no information about any of the existing
complementary observables is available.

\textit{Generating informationally complete measurements.}  
As we will now show, this interpretation no longer holds when only a
subspace $\mathcal{S}$ of $\mathcal{H}$ is considered. Let us specify
$\mathcal S$ by introducing the projector 
$
P_S= \sum_{s=0}^S  |s \rangle \langle s| \, ,
$
where $|s\rangle$ are eigenstates of $P_S$ and $S$ is the dimension.
By projecting  the original measurement on $\mathcal S$, a POVM is
induced in this subspace, namely
\begin{equation}
  \label{induced} 
  \sum_k \Pi_k = 
  \sum_k P_S \, |k \rangle \langle k |\,  P_S = \openone_S \, ,
\end{equation}
whose elements, in general, no longer commute $[\Pi_{k},\Pi_{k'}]
\neq 0$. Indeed, since the original commuting projections 
have different overlaps with the subspace $\mathcal{S}$, 
their mutual properties (commutators) are not preserved.
In this way, an informationally complete POVM may be
generated. Obviously, this observation has many potential 
applications beyond tomography, although, due to strict
space limitation only that topic will be \sugg{discussed.}

The protocol we propose consists of the following steps: (i) A
reconstruction subspace $\mathcal{S}$ is selected according to the
particular experiment, in such a way that all the relevant states are
included. (ii) An experimentally feasible von Neumann projection is
chosen.  (iii) The effective POVM induced in $\mathcal{S}$, as given
by Eq.~\eqref{induced}, is calculated and its informational
completeness is checked with the help of condition \eqref{condition}.
If the induced POVM is informationally complete, the task is finished,
otherwise the whole procedure is repeated with different choices of
either the von Neumann projection or the reconstruction subspace or
both.

Before we proceed further, let us comment on the differences between
our protocol and the Naimark extension~\cite{naimark}, which is
another way of representing POVMs by projective measurements.  This
extension works by enlarging the Hilbert space with an ancilla, so the
projective measurement acts on the product space of the system and
ancilla. In our approach, the possibility of representing a
tomographical scheme by a projective measurement stems from the
available prior information.  In fact, the unpopulated states or
unused range of variables play the role of ancilla here and,
consequently, the measurement acts on a sum rather than a product
space.

\textit{Optical vortices.}
As a relevant example, we use our protocol for the tomography of
optical vortices. As the wave function (or density matrix) in
quantum theory, any transverse distribution of complex amplitude (or
coherence matrix) can be decomposed in a complete basis; the
Laguerre-Gauss modes being a very convenient one
\begin{equation}
  \label{LGbeams}
  \LG{\ell}{p} (x,y) = \langle x,y |\ell, p \rangle \propto 
  r^{|\ell|} L_p^{|\ell|} (2r^2) e^{-r^2} e^{i \ell \phi} \, ,
\end{equation}
where $r^2=x^2+y^2$ and $\phi=\arctan(y/x)$ are polar coordinates in
the transverse plane and $L_p^\ell$ is a generalized Laguerre
polynomial. It is well known~\cite{LG} that $\LG{\ell}{p}$ beams
exhibit helicoidal wavefronts that induce a vortex structure and carry
orbital angular momentum of $ \hbar \ell$ per photon. Suppose a photon
has been emitted into a superposition of modes, and we need to identify
the resulting state. In general, this is an involved
task~\cite{zeilinger,white,calvo} requiring the use of complicated
optical devices. However, provided that only beams with bounded
vorticities (i.e., values of $|\ell|$) are considered, as it is usually
the case, our protocol can be employed and an informationally complete
measurement can be generated from a very basic one, such as a single 
transverse intensity scan that is easy to record.  In the language of 
quantum theory, this intensity scan is just
$
  I(x,y) \propto \mathrm{Tr}(\varrho|x,y \rangle \langle x,y|),
$
where $x$ and $y$ denote now the coordinates of a given pixel of the
position-sensitive detector. Although detections in any pair of pixels
are always compatible, in a subspace with bounded vorticities
\sugg{noncommuting} POVM elements can be induced.
\begin{figure}
  \includegraphics[width=0.7\columnwidth]{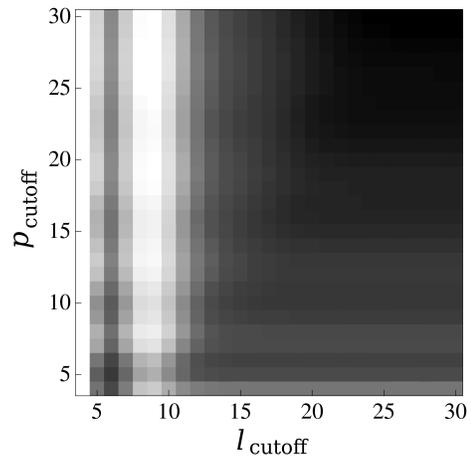}
  \caption{Incompatibility \sugg{(computed as the norm of commutator)} 
    of the detections at two spatially separated pixels of a CCD 
    camera in a truncated Hilbert space $p=0,\ldots,p_{\mathrm{cutoff}}$, 
   $\ell=-\ell_{\mathrm{cutoff}},\ldots,\ell_{\mathrm{cutoff}}$.
    Black (white) color means compatible (strongly
    incompatible), respectively.}
\label{fig:comut}
\end{figure}
This is illustrated in Fig.~\ref{fig:comut}, which shows the
\sugg{noncommutativity} (incompatibility)
corresponding to the positions $(x,y) = (0, 0)$, and $(0,1)$ [in the
same units of Eq.~\eqref{LGbeams}].  Truncating the Hilbert space at
smaller vorticities typically leads to stronger \sugg{noncommutativity},
although some nonmonotonicity is also observed as oscillations 
of gray shades \sugg{appearing} from the top-right to the bottom-left corner.

\begin{figure}
\centerline{\includegraphics[width=\columnwidth]{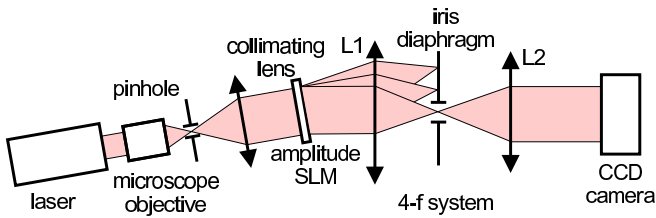}}
\caption{
Experimental setup of vortex tomography by means of compatible observations.
\label{fig:setup}}
\end{figure}

\textit{Experiment.}
To demonstrate the potential of the procedure, a full tomography of an
optical vortex field from a single intensity scan has been performed
in a controlled experiment.  The experimental scheme is shown in
Fig.~\ref{fig:setup}.  The beam generated by a He-Ne laser is
spatially filtered by a microscope objective and a pinhole. After the
beam is expanded and collimated by a lens, it impinges on an amplitude
spatial light modulator (CRL Opto, $1024 \times 768$ pixels)
displaying a hologram computed as an interference pattern of the
required light and the inclined reference plane wave.

Light behind the hologram consists of three diffraction orders $(-1,
0, +1)$, which can be separated and Fourier filtered by means of the
4$f$ optical system consisting of the lenses L$_{1}$ and L$_{2}$, and an
iris diaphragm. The undesired 0th and -1st orders are removed by an
aperture placed at the back focal plane of the lens L$_{1}$. This
completes the preparation of a given state of light.

Finally, a collimated beam with the required complex amplitude profile
is obtained at the back focal plane of the second Fourier lens
L$_{2}$, where a transverse intensity scan $I(x,y)$ is acquired by a
CCD camera.  In the image plane, each pixel detection can be approximated
by \sugg{a projection} on the position eigenstates $|x,y \rangle \langle
x,y|$. As it has been shown above, while such detections are
compatible in the full infinite-dimensional Hilbert space, an
informationally complete POVM is induced in a subspace of truncated
vorticities.

\begin{figure}
  \centerline{\includegraphics[width=0.33\columnwidth]{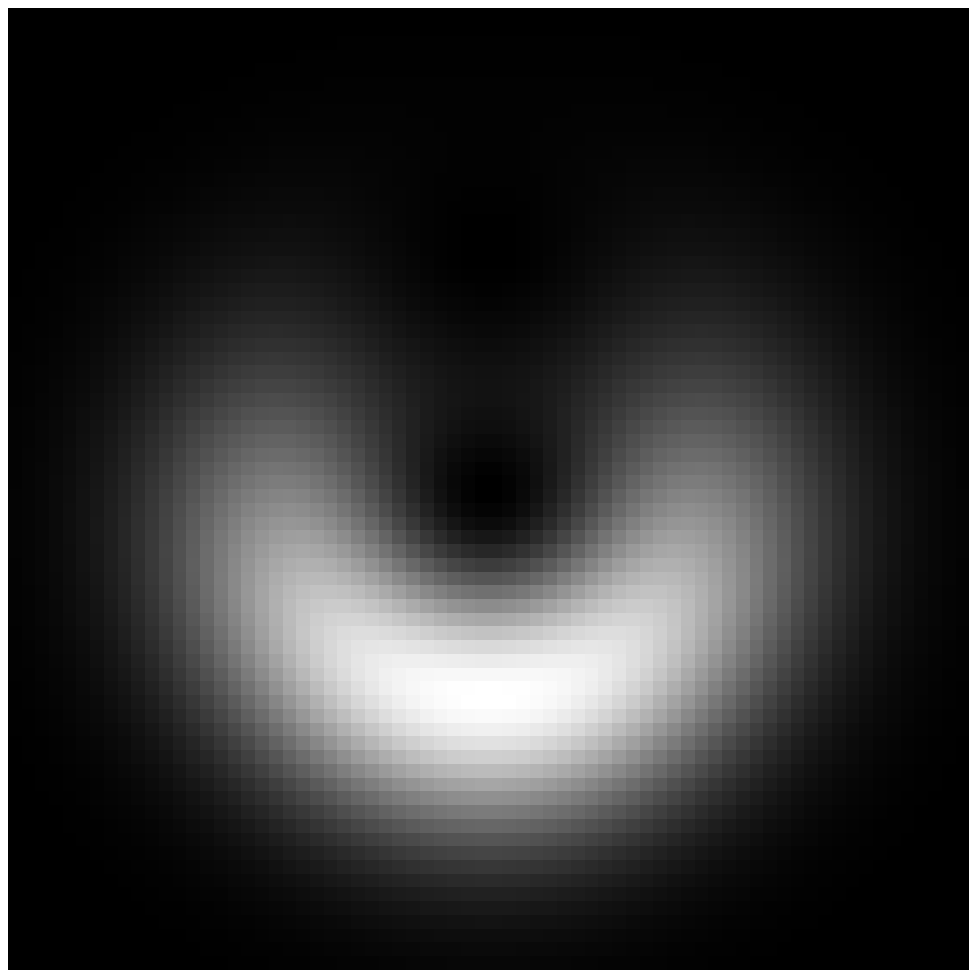}
    \includegraphics[width=0.33\columnwidth]{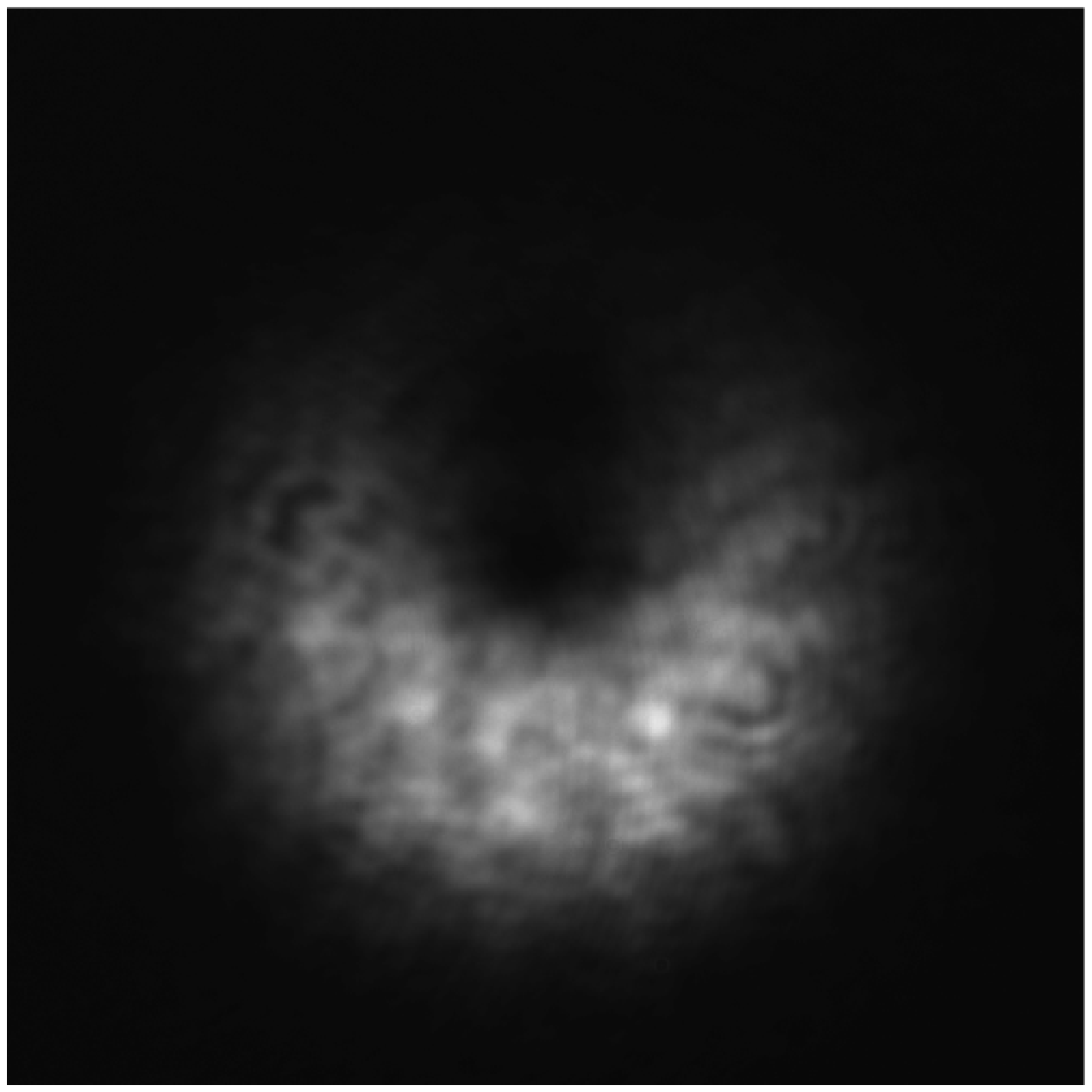}
    \includegraphics[width=0.33\columnwidth]{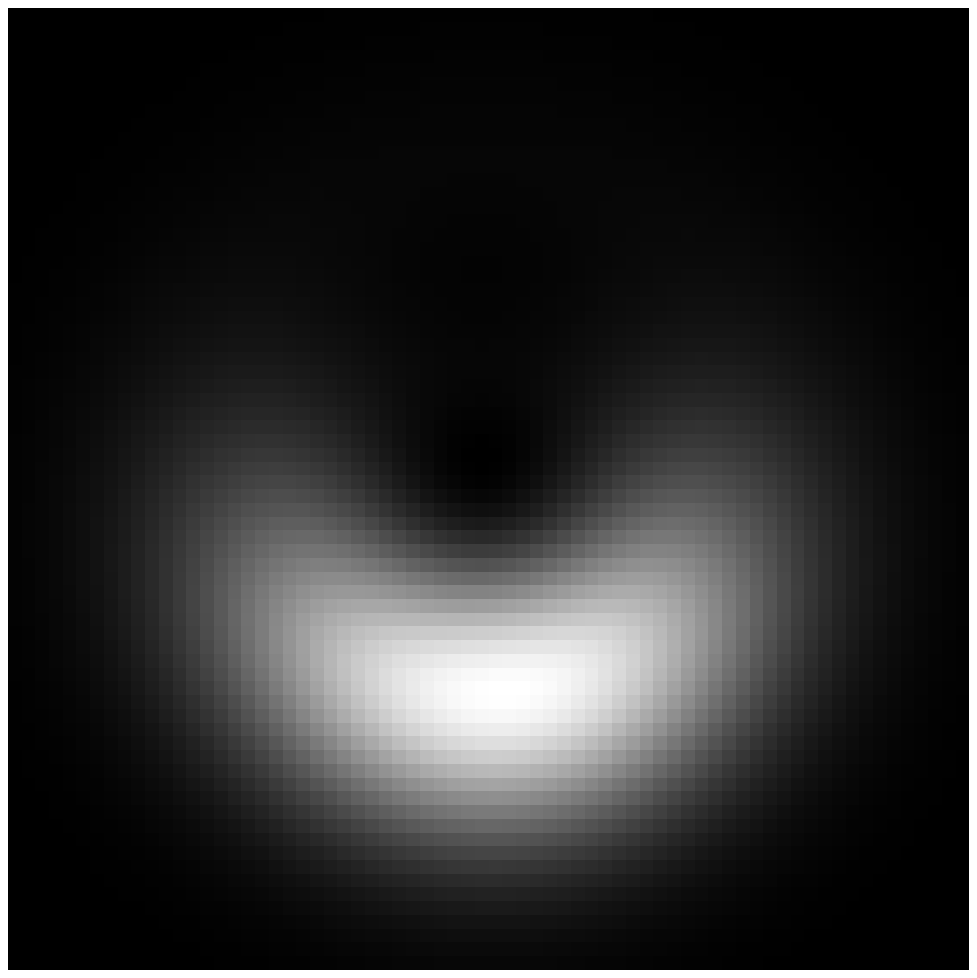}}
  \caption{\label{fig:expintens} Experimental tomography of optical
    vortex fields. From left: ideal intensity distribution, measured
    intensity distribution, and the corresponding best theoretical fit
    of measured data.}
\end{figure}

\begin{figure}[b]
  \centerline{\includegraphics[width=0.5\columnwidth]{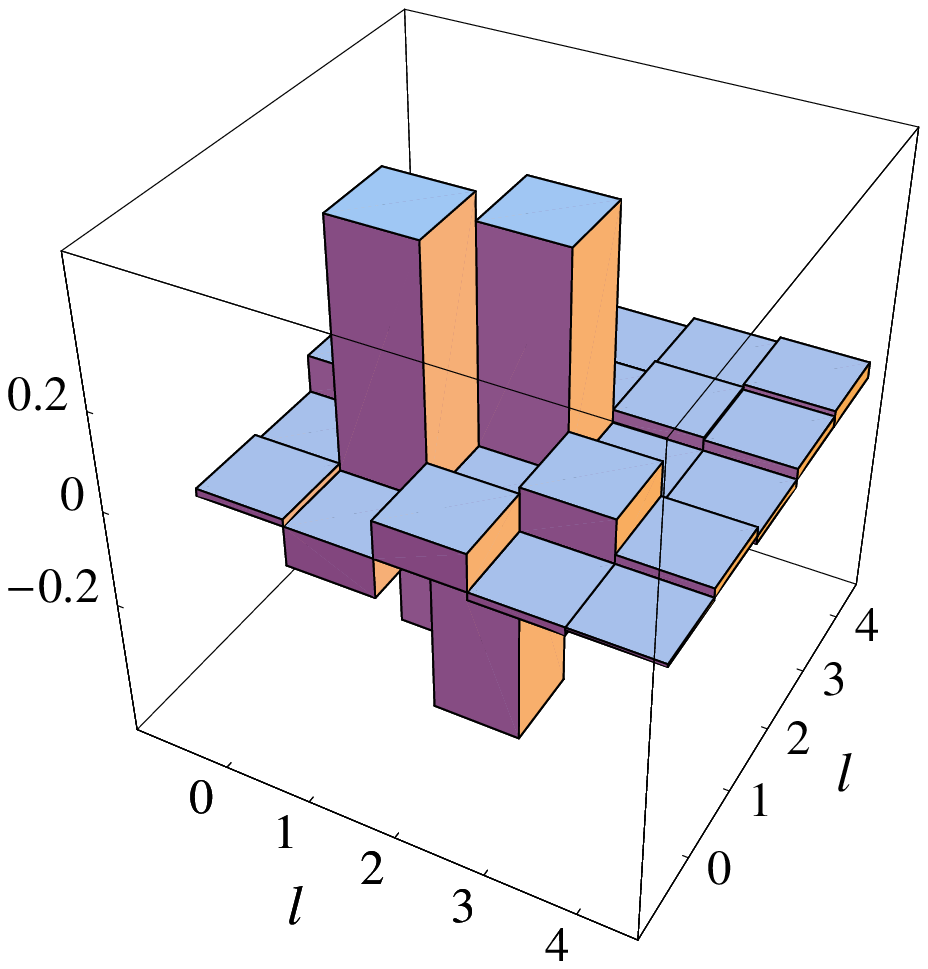}
    \includegraphics[width=0.5\columnwidth]{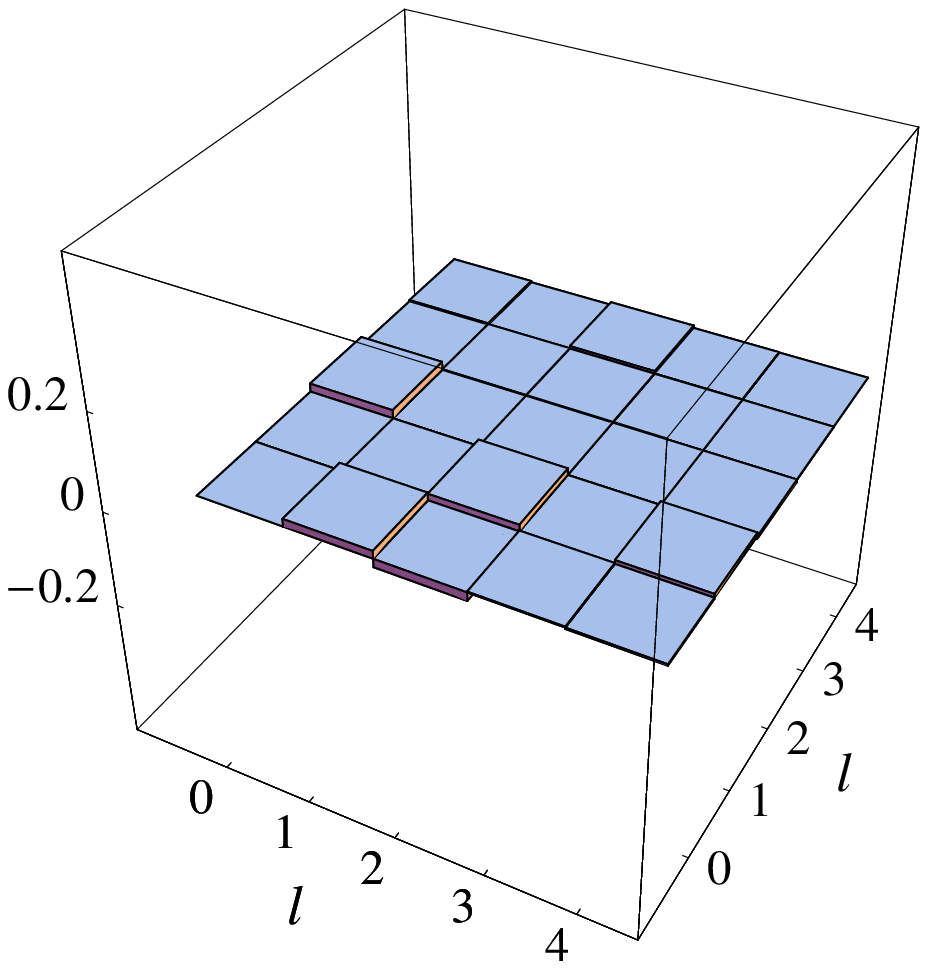}}
  \caption{\label{fig:exprecon} Real (on the left) and imaginary (on
    the right) elements of the reconstructed density matrix.}
\end{figure}

In our experiment the superposition
\begin{equation}
  \label{truestate}
  |\Psi \rangle = \frac{1}{\sqrt{2}} 
  (|\ell=1, p=0 \rangle + | \ell=2, p=0 \rangle)
\end{equation}
was prepared by letting an amplitude spatial light modulator to
display an interference pattern of the transverse amplitude
\sugg{$\langle x,y|\Psi\rangle$} and a reference plane wave, as
mentioned above.  Results for this state are shown in
Fig.~\ref{fig:expintens}.  The ideal intensity distribution in the
detection plane $I(x,y) \propto |\langle x,y|\Psi\rangle|^2$ is shown
in the left panel.  This should be compared to the corresponding noisy
recorded image shown in the middle panel. Finally, the right panel
shows the best fit obtained with a maximum-likelihood
algorithm~\cite{prl} in the subspace $p=0$ and $ \ell= 0,\ldots,4$.
The reconstructed $5$-dimensional density matrix is shown in
Fig.~\ref{fig:exprecon}.  Notice that, due to experimental
imperfections \sugg{ (such as a discrete structure of the spatial
light modulator, detection noise, etc.)}, the reconstructed state
\sugg{slightly differs} from the ideal one (typical fidelities in our
experiment are $F \approx 96\%$).  \sugg{In view of the complexity of
the system and the simplicity of the experiment, we consider this to
be a very good result}.

\begin{figure}
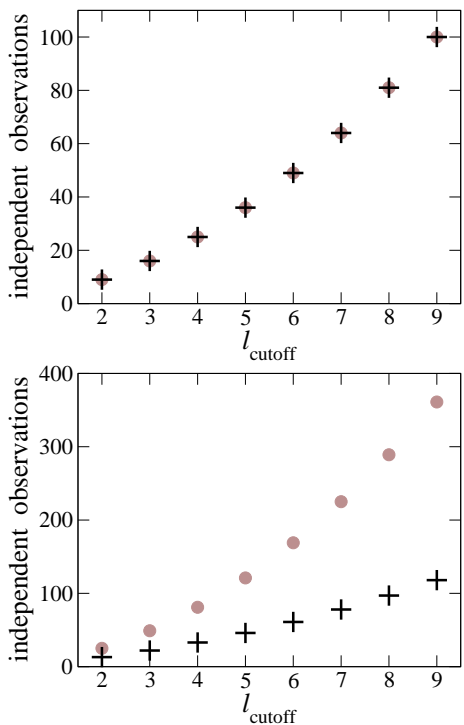

\includegraphics[width=0.7\columnwidth]{rank2a}
\includegraphics[width=0.7\columnwidth]{rank1a}
\caption{\label{fig:rank} Informational completeness of measurements
on vortex beams generated  by a CCD camera with $11\times 11$ pixels. 
The number of independent measurements $\Pi_k$ generated from those $121$ CCD 
detections are shown by circles for different truncations of the Hilbert space $P_S$.
The number of independent measurements required for a complete tomography
in the same reconstruction subspace is indicated by crosses.
The reconstruction subspaces are truncated as follows.  
Upper panel: $p=0$, $\ell=0,\ldots, \ell_{\mathrm{cutoff}}$; 
bottom panel: $p=0$, $\ell=-\ell_{\mathrm{cutoff}}, \ldots, \ell_{\mathrm{cutoff}}$.
}
\end{figure}

Given the promising performance of the proposed scheme in this
\sugg{proof-of-principle} experiment, the natural question is whether
an experimentally feasible von Neumann measurement (such as a single
intensity scan by a CCD camera with possibly very fine resolution)
would furnish an informationally complete measurement for any
reconstruction subspace.  To get some insights into this problem, we
consider two different scenarios related to the experiment above (see
Fig.~\ref{fig:rank}).  In the first \sugg{case,} only photons with
nonnegative vorticities are considered: the full tomography from a
single intensity scan is always possible.  In the second case, both
positive and negative vorticities are allowed.  Here a single
intensity scan fails to provide complete information. It is easy to
see why: since the intensity profiles of the Laguerre-Gauss modes
$\LG{\ell}{p}$ and $\LG{-\ell}{p}$ are the same, perfect
discrimination between states with positive and negative vorticities
is not possible. Interestingly enough, \textit{some} information about
the negative part of the angular momentum spectrum is still available
(see, e. g., the crosses in the plots for the same truncation
$\ell_{\mathrm{cutoff}}$), \sugg{as it is also obvious} from the fact
that \sugg{the phases} $\exp(i\ell\phi)$ and $\exp(-i\ell\phi)$ in
superpositions like $\LG{\ell}{0}+\LG{1}{0}$ and
$\LG{-\ell}{0}+\LG{1}{0}$ can be distinguished via interference with
the other mode. This partial information is however not sufficient for
the full characterization of this part of the reconstruction subspace.
Provided one wants to keep the simple intensity detection, it is
always possible to use a fixed unitary transformation prior to
detection to optimize the scheme.  For instance, by increasing angular
momentum of the measured beam by $\ell_{\mathrm{cutoff}}$ (using,
e. g., a charged fork-like hologram) the reconstruction subspace can
be moved inside the nonnegative part of the angular momentum spectrum.
This example nicely illustrates the role of prior information in
experimental quantum tomography.

\textit{Conclusions.}  
We have shown that simple compatible observations may provide full
information about the measured system when some prior information is
available. This prior information does not only bring about a
quantitative improvement of our knowledge, but may also make feasible
a no-go task.  Based on this observation, an efficient protocol was
sketched providing the full characterization of complex systems from
simple measurements.  This was demonstrated in an experiment with
photonic vortices. In our opinion, \sugg{this constitutes an improvement
that will have a significant benefit in the number of different physical
architectures where quantum information experiments are being performed.}

This work was supported by the Czech Ministry of Education, Projects
MSM6198959213 and LC06007, the Spanish Research Directorate, Grants
FIS2005-06714 and FIS2008-04356.

\end{document}